\documentclass[10pt]{amsart}
\usepackage{amsfonts}
\usepackage{epsfig}
\usepackage{latexsym}
\usepackage{amsmath,amssymb,amsfonts,amsthm,graphics}
\usepackage{eucal}
\usepackage{eufrak}
\usepackage[all]{xypic}
\usepackage{xspace}

\theoremstyle{plain}
\newtheorem{theorem}{Theorem}

\newtheorem{lemma}{Lemma}

\theoremstyle{definition}

\theoremstyle{remark}

\numberwithin{equation}{section}

\begin{document}

\title[Neutral Networks of Sequence to Shape Maps]
      {Neutral Networks of Sequence to Shape Maps}
\author{Emma Y. Jin, Jing Qin and Christian M. Reidys$^{\,\star}$}
\address{Center for Combinatorics, LPMC-TJKLC \\
         Nankai University  \\
         Tianjin 300071\\
         P.R.~China\\
         Phone: *86-22-2350-5133-8013\\
         Fax:   *86-22-2350-9272}
\email{reidys@nankai.edu.cn}
\thanks{}
\keywords{combinatory map, component, diameter, neutral network, shape,
          bipartite}
\date{June, 2007}
\begin{abstract}
In this paper we present a combinatorial model of sequence to shape maps. 
Our particular construction arises in the context of representing nucleotide 
interactions beyond Watson-Crick base pairs and its key feature is to 
replace sterical by combinatorial constraints. We show that these combinatory 
maps produce exponentially many shapes and induce sets of sequences which 
contain extended connected sub graphs of diameter $n$, i.e.~we show that 
exponentially many shapes have neutral networks.
\end{abstract}
\maketitle
{{\small
}}

\section{Introduction}


\subsection{Background}
\begin{figure}[ht]
\centerline{%
\epsfig{file=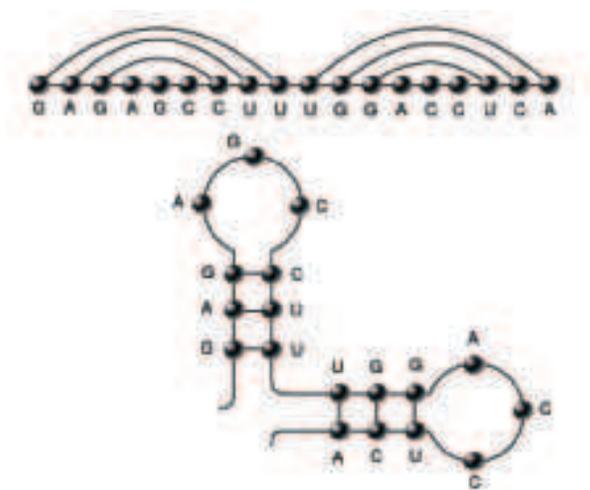,width=0.8\textwidth}\hskip15pt }
\caption{\small RNA secondary structures. Diagram representation
(top): the primary sequence, {\bf GAGAGCCUUUGGACCUCA}, is drawn
horizontally and its backbone bonds are ignored. All bonds are drawn
in the upper half plane and secondary structures have the property
that no two arcs intersect and all arcs have minimum length $2$.
Outer planar graph representation (bottom). } \label{F:1}
\end{figure}

Arguably one of the greatest challenges in present day biophysics is the
understanding of sequence structure relations of bio polymers. For one
particular class of bio polymers, the ribonucleic acid (RNA) secondary
structures, (Fig.~\ref{F:1}) molecular folding maps have been systematically 
analyzed by Schuster~{\it et.al.} \cite{Fontana:98,Schuster:94,Schuster:02}.
Folding maps play a central role in understanding the evolution of molecular
sequences. Specific properties like, for instance {\it shape space covering}
\cite{Schuster:95B} and {\it neutral networks} (Fig.~\ref{F:2})
\cite{Reidys:97a} are critical
for what may be paraphrased as ``molecular computation by white noise''.
For instance, neutral networks played a central role in the {\it Science}
publication authored by E.~Schultes and P.~Bartels {\it One sequence, two
ribozymes: implications for the emergence of new ribozyme folds}, (v289, n5478,
448-452) where the authors designed experimentally a single RNA sequence (whose
existence is implied by the intersection theorem in \cite{Reidys:97a}) that
folds into two different, non-related, RNA secondary structures 
\cite{Clote:05}.
Exhaustive enumeration of sequence spaces and subsequent detailed analysis
of the mappings for {\bf G},{\bf C}-sequences of length $30$ were undertaken
in \cite{Gruener:95a,Gruener:95b}. In addition detailed analysis of
neutral networks as well as exhaustive enumeration of 
{\bf G},{\bf C},{\bf A},{\bf U}-sequences can be found in \cite{Goebel:04}.
The findings were intriguing. Folding maps into RNA secondary structures
exhibit a collection of distinct properties which makes them ideally suited
for evolutionary optimization.\\
{\sf (a)} Many structures have preimages of sequences (neutral networks)
          which have large components and large diameter.\\
{\sf (b)} Many structures have the property that any two of them have neutral
          networks that come close in sequence space.\\
Obviously, {\sf (a)} is of central importance in the context of
neutral evolution. Since replication is erroneous and only few if
not single nucleotides can be exchanged the preimages of structures
must contain large connected components.
\begin{figure}[ht]
\centerline{%
\epsfig{file=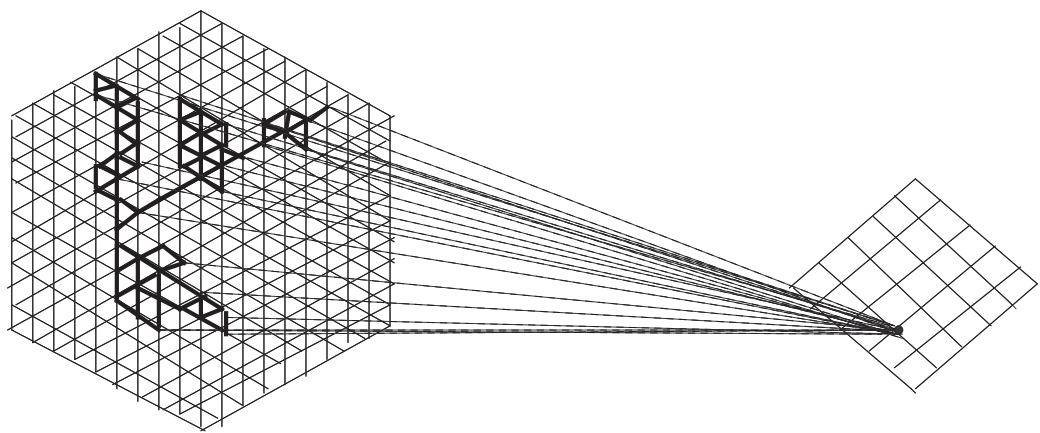,width=0.95\textwidth}\hskip15pt } \caption{\small
The neutral network of a structure. Sequence space (right) and shape 
space (left) represented as lattices.
We draw the edges between two sequences bold if they map into the one 
particular structure on the left. The two key properties of neutral nets
are their connectivity and percolation. They allow sequences to move
while maintaining a shape through sequence space.} \label{F:2}
\end{figure}
{\sf (b)} showed that (many) new structures can easily be found
during a random walk on a neutral network using only steps in which
a single nucleotide is altered (point mutations). 

Folding maps, however, are not obtained analytically. They are a result of a
computer algorithm, based on the combinatorial analysis of RNA
secondary structures pioneered by Waterman~{\it et.al.}
\cite{Waterman:94,Waterman:78a,Waterman:79}. It has to be remarked
in this context that comparative sequence analysis \cite{Woese:93,Puglisi:99} 
provides more reliable means for determining the secondary structure 
of biological RNA \cite{Doudna:99}, i.e.~folding 
maps represent already an abstraction.
In order to step beyond the secondary structure paradigm two main approaches 
with distinct goals are: (1) to study more advanced nucleotide interactions
in RNA, like for instance pseudoknots, base triples or (2) consider genuine
abstractions of molecular structures not aiming to model a biophysical 
folding map. In \cite{Reidys:07rna1} we
pursue the first by developing the combinatorics of RNA structures
with pseudoknots and in this contribution the second by studying
combinatory maps. While (1) eventually produces the mathematical
framework enabling us to derive more advanced representations (which
eventually result in folding algorithms capable of producing
structures like phenylalanine tRNA) (2) provides insights on the
core question of which principles produce sequence to structure maps
suitable for evolution. A type (2) abstraction inevitably evokes 
skepticism since what can possibly be gained if no attempt is made 
to mimic the biological reality? However, we argue that sometimes it 
is exactly the right strategy to fundamentally understand the object under 
investigation.

\subsection{Structures and correlations}
A well studied class of maps over sequence spaces are the
NK-landscapes introduced by Kauffman \cite{Kauffman:93}, where each
index (locus) of a binary $n$-tuple viewed as the genotype composed
by $n$ loci is randomly linked to K other indices. The idea is that
a locus $i$ makes a contribution to the total fitness of the
genotype which depends on the value of the allele ($0$ or $1$) at
$i$ and the values at each of the epistatically linked loci. To each
of those $2^{{\rm K}+1}$ combinations there is a value (fitness)
assigned uniformly at random. The apparent lack of neutrality led
Barnett \cite{Barnett:98} to refine NK-landscapes by NKp-landscapes,
introducing a probability $p$ with which an arbitrarily chosen
allelic combination makes no contribution to the fitness. Our
approach is connected to Kauffmann's intuition in that we consider a
molecular structure as a combinatorial representation of
nucleotide-correlations. As for nucleotide-correlations 
observations {\sf (a)} and {\sf (b)} are not bound to the particular
concept of RNA secondary structures. For instance Stadler {\it et.al.}
\cite{SchusterStadler:99}
as well as Bastolla {\it et.al.} have shown \cite{Porto:03} that neutral 
networks exist for proteins, where nucleotide interactions are much more 
involved \cite{Reidys:00p1}. 
Therefore it is certainly not the uniqueness of Watson-Crick base pairings
implying the existence of neutral networks. Our particular approach
comes from this correlation perspective and observations from molecular 
interaction in RNA molecules.
\begin{figure}[ht]
\centerline{%
\epsfig{file=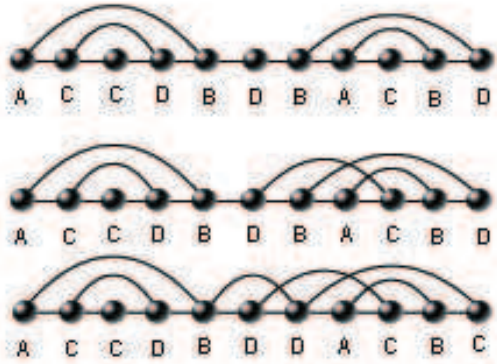,width=0.8\textwidth}\hskip15pt } \caption{\small
Beyond secondary structures. Suppose we are given an abtract alphabet 
$\{{\bf A},{\bf B},{\bf C},{\bf D}\}$ with base pairs 
$\{\{{\bf A},{\bf B}\},\{{\bf D},{\bf C}\},\{{\bf D},{\bf B}\}\}$.
We present diagram representations of a secondary structure (top),
$3$-noncrossing structure (middle) and a $2$-diagram structure (bottom).
The difference between the first two structures is the crossing of bonds 
and the difference between the second two is the number of interactions
for a nucleotide.} 
\label{F:3}
\end{figure}
First there are secondary and tertiary interactions \cite{Doudna:99},
the latter typically involving secondary structural elements.
Furthermore interaction within RNA molecules can be categorized into
three classes, helix-helix interaction, loop/bulge-helix and
loop-loop interaction \cite{Westhof:92a,Doudna:99}. The structure of
phenylalanine tRNA, and the hammerhead ribozyme \cite{Wedekind:98}
have served as paradigms in this context. Base
triples and tetra-loops, as well as pseudoknots,
\cite{Westhof:92a,Konings:95a,Chamorro:91a,Science:05a}
representing loop-loop interactions have led to generalizations of
the secondary structure concept. These interactions are subject to
steric constraints arising from the biochemistry of the interactions 
involved. 
These observations give rise to two different combinatorial abstractions:
the consideration of $k$-noncrossing chemical bonds and of
$2$-diagrams i.e.~a graph whose vertices are drawn as a horizontal
line having degree less than two (and the combination of them,
$k$-noncrossing $2$-diagrams). The notion of $k$-noncrossing arises
naturally in the context of pseudoknots leading to the concepts of
$k$-noncrossing RNA structures \cite{Reidys:07rna1} and to Stadler's
bi-secondary structures \cite{Stadler:99} (which are exactly the
planar $3$-noncrossing RNA structures). The notion of $2$-diagrams
comes up when restricting nucleotide interactions to at most two and
therefore allowing the expression or interactions of secondary
structure elements.

\section{The Basic Construction}
The notion of $2$-diagrams discussed in the introduction is exactly the 
motivation of our
particular approach. In the following we detail how to derive molecular
shapes in which each nucleotide has at most two interactions but which, in
difference to biophysical structures, have combinatorial constraints 
on their nucleotide interactions.
This idea is to the best of our knowledge new.
For a given alphabet base pairing rules specify which nucleotides
can pair. However, not any two nucleotides are able to establish a
bond. For instance, they may be restricted by conditions like no two
edges can cross each other when representing a shape as a diagram
\cite{Stadler:99}. The non-crossing condition and uniqueness of base
pairs are two key properties of RNA secondary structures and allow
for Motzkin-path enumeration and tree bijections
\cite{Waterman:94,Waterman:79,Zuker:79,Waterman:78a,Schuster:98}. We
replace these restrictions on nucleotide interactions by stipulating
that {\sf (a)} there exists some base graph $H$ whose sole purpose
is to restrict all possible correlations and {\sf (b)} we are given
a symmetric relation $\mathcal{R}$, tantamount to a base pairing
rule. In order to avoid any confusion we work over the abstract
alphabet $\{{\bf A},{\bf B},{\bf C},{\bf D}\}$. 

\begin{figure}[ht]
\centerline{%
\epsfig{file=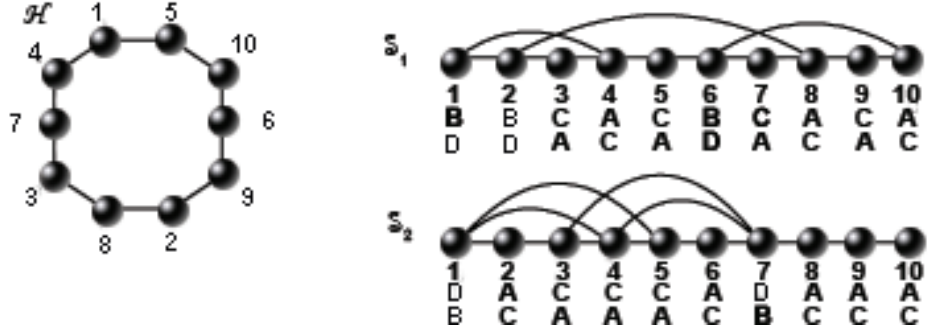,width=1.0\textwidth}\hskip20pt
}
\caption{\small Combinatory maps: the base graph $\mathcal{H}$ is displayed
on the l.h.s.. The r.h.s.~shows two shapes $\mathcal{S}_1$ and
$\mathcal{S}_2$ with two particular sequences that are contained in
their respective preimages. For both sequences the shapes are
maximal, i.e.~not a single $\mathcal{H}$-edge can be drawn without
violating base pairing rules, here $\{\{{\bf A},{\bf B}\}$, $\{{\bf D},
{\bf C}\}$, $\{{\bf D},{\bf B}\}\}$.}
\label{fig:base}
\end{figure}
In this framework a shape $\mathcal{S}$ of a sequence is then the unique 
maximal $H$-subgraph subject to the property that for any 
$\mathcal{S}$-edge the incident nucleotides satisfy $\mathcal{R}$. 
It is remarkable that this simple definition already produces a well 
defined sequence to structure map! Moreover this definition is in line
with the biological point of view: mapping sequences into shapes 
rather than fixing some shape and then to consider its sequences. 
It now can be asked what the right choice of $H$ should be and how robust 
the respective conclusion are. As for dependency on $H$ the answer is that 
it a.s.~(almost surely in the sense of random graph theory, i.e.~in the 
limit of long sequences) depends on the number of edges, only. 
Therefore, the choice of $H=\mathcal{H}$ is not critical for the validity of
the main results. To understand why, we consider a generalization of the 
concept of combinatory maps, i.e.~combinatory maps induced the random graph 
$G_{n,p}$ (the random graph in which each edge is selected with independent 
probability $p$). In the sub-critical phase these random combinatory maps 
a.s.~produce, modulo constants, all properties of the maps induced by 
$\mathcal{H}$ (Theorem~\ref{T:nn}). 

{\bf Theorem}\cite{Reidys:07priv} {\bf (Neutral networks)} {\it
Let $p_n=\frac{1-\epsilon}{n}$, $\beta<\sqrt{2}$ and
suppose $\omega_n$ tends to $\infty$
arbitrarily slowly and $\vartheta_{G_{n,p}}$ is a random combinatory map.
Then there exist with high probability at least $\beta^n$ shapes $\mathcal{S}$
with the following two properties:\\
{\sf (I)} the set of all sequences mapping into $\mathcal{S}$ has a connected
          component of size at least $\left(\sqrt{2}\right)^{n}$\\
{\sf (II)} the set of all sequences mapping into $\mathcal{S}$ percolates, i.e.
           has diameter $n-\omega_n$.\\
}

The great advantage of choosing $H=\mathcal{H}$ is the simplicity and
algorithmic nature of all proofs. We can explicitly construct all paths
involved by diagram chasing. In contrast, the proof of the above
result is based on a non trivial analysis of tree components in the 
random graph $G_{n,p}$.
We have
the following situation: let $H$ be a graph over $\{1,2,\ldots,n\}$,
$\mathcal{A}= \{{\bf A},{\bf B},{\bf D}, {\bf C}\}$ and $Q_4^n$ be
the generalized $n$-cube, $Q_4^n$, i.e.~the graph over the sequences
$(x_1,\dots,x_n)$, where $x_i\in\mathcal{A}$ and in which two
sequences are adjacent if they differ in exactly one nucleotide. Let
$d(v,v')$ be the number of nucleotide by which $v$ and $v'$ differ.
A component of a graph $H$ is a maximal connected subgraph. We
consider relations $\mathcal{R}$ over the abstract alphabet
$\mathcal{A}=\{{\bf A},{\bf B},{\bf D}, {\bf C}\}$,
i.e.~$\mathcal{R}\subset \mathcal{A}\times \mathcal{A}$ satisfying
the following three conditions
\begin{eqnarray}
\label{E:1} (x,y)\in \mathcal{R}  & \Leftrightarrow &  (y,x)\in \mathcal{R} \\
\label{E:2} (x,y)\in \mathcal{R}  & \Rightarrow &\ x\neq y        \\
\label{E:3} \forall\, x\neq z\quad (x,y)\in
\mathcal{R}\,\wedge\,(y,z)\in \mathcal{R} & \Rightarrow &
(x,z)\not\in \mathcal{R} \ .
\end{eqnarray}
These conditions are motivated from abstracting form $2$-D and $3$-D
interactions of the phenylalanine tRNA and the hammerhead ribozyme
\cite{Doudna:99}. In both molecules mutual interactions of $3$-nucleotides
are absent but multiple pair interactions are responsible for the tertiary
structure. In view of eq.~(\ref{E:1}) and eq.~(\ref{E:2}) each relation can be
viewed as a graph over $\{{\bf A},{\bf B},{\bf D},{\bf C}\}$ and obviously,
eq.~(\ref{E:3}) is equivalent to this graph being 
bipartite\footnote{For instance, it is easy to check that the relation
implied by all Watson-Crick base pairs (i.e.~\{({\bf A},{\bf U}),({\bf
U},{\bf A}),({\bf G},{\bf C}),({\bf C},{\bf G})\}) and \{({\bf G},{\bf
U}),({\bf U},{\bf G})\}, satisfy
conditions eq.~(\ref{E:1}), eq.~(\ref{E:2}) and eq.~(\ref{E:3}).}.
We will be particularly interested in the base pairing rule
$\mathcal{R}^\dagger$ represented as the graph $\diagram {\bf A}\rline
&{\bf B} \rline & {\bf D}\rline & {\bf C}
\enddiagram$ i.e.~we allow for the following interactions:
$\{\{{\bf A},{\bf B}\},\{{\bf D},{\bf C}\},\{{\bf D},{\bf B}\}\}$.
In this sense our nucleotide interactions are more general than those 
of RNA secondary structures since, for instance, we can express coaxial
stacking of helical regions and the formation of isosteric
${\bf C}\cdot {\bf G}-{\bf G}$ triples \cite{Doudna:99}.
We introduce the $H$-subgraph $H_{\mathcal{R}}(v)$ having vertex and edge set
given by
\begin{equation}
V_{H_{\mathcal{R}}(v)}=\{1,\dots,n\}, \quad \text{\rm and}\quad
E_{H_{\mathcal{R}}(v)}=\{\{i,k\} \mid \{i,k\}\,\text{\rm is an
$H$-edge and}\,
          (x_i,x_k)\in \mathcal{R}\}
\end{equation}
and call $H_{\mathcal{R}}(v)$ a shape $\mathcal{S}$ and the mapping
$
\vartheta_{H}:Q_4^n \longrightarrow \{\mathcal{S}\mid
\mathcal{S}=H_{\mathcal{R}}(v)\}
$
a combinatory map. Note that the above construction entails an implicit
notion of maximality, i.e.~ a shape of a sequence $(x_1,\dots,x_n)$ is the
maximal $H$-subgraph which satisfies $\mathcal{R}^\dagger$ for all
$2$-sets of coordinates $\{x_i,x_j\}$, $\{i,j\}$ being a $H$-edge.
In this sense a shape represents a saturated structure.
As for $\mathcal{H}$, suppose first $n$ is even. We set
$C_{n}(1)$ to be the graph over $\{1,\dots,n\}$ with edge set
$\{i,i+1\}$ where the vertices are labeled modulo $n$.
Let $\sigma_n$ some permutation of $n$-letters, we then set
$C_{n}(\sigma_{n})$ with edges $\{\sigma_n(i),\sigma_n(i+1)\}$
and $\mathcal{H}=C_{n}(\sigma_n)$. Next assume $n$ is odd. Then we
select an arbitrary element of $\{1,\dots,n\}$, say $u$ and define
$\mathcal{H}=C_{n-1}(\sigma_{n-1})\cup \{u\}$ i.e.~the graph
with edges $\{\sigma_{n-1}(i),\sigma_{n-1}(i+1)\}$ for $i\neq u$ and
$i+1\neq u$, where $\sigma_{n-1}$ is an arbitrary permutation of
$\{1,\dots,n\}\setminus \{u\}$. To summarize we have
\begin{equation}\label{E:H}
\mathcal{H}=
\begin{cases}
C_n(\sigma_n)                   & \ \text{\rm for $n$ even}\\
C_{n-1}(\sigma_{n-1})\cup \{u\} &\  \text{\rm for $n$ odd} \ .
\end{cases}
\end{equation}

\section{Shapes}

In this section we answer the following basic questions: \\
{\sf (1)} What is the relation between base pairing rules and the 
resulting molecular shapes? \\
{\sf (2)} How many shapes does a combinatory map have?\\
{\sf (3)} Are there ``many'' shapes with large sets of sequences folding 
           into them?\\
All of the above properties are central for RNA secondary structures and
none of them can be answered analytically, despite the fact that we have
generating functions for RNA secondary structures.
For instance, it is impossible to
assess {\it a priori} how many secondary structures have an actual sequence
folding into them. The number of RNA structures that actually occur as minimum
free energy structures can be much smaller than the total number. For $n=16$,
due to finite size effects for the RNA folding, only $63\%$ of the possible
RNA structures are realized as minimum free energy structures 
\cite{Goebel:04}. 

Let us begin by providing some more background: graph $H'$ is
called an induced subgraph of $H$ iff there exists some set
$M\subset \{1,\dots,n\}$ such that $E_{H'}=\{\{i,j\}\mid \{i,j\}\in
E_H\,\wedge i,j\in M\}$. Intuitively, induced subgraphs come from
vertex sets and are far more restricted that arbitrary subgraphs. We
now give a simple example of the fact that not every bipartite
subgraph of a shape is a shape. For instance, consider
$\vartheta_H:Q_4^6\longrightarrow \{H'<H\}$ where
\begin{equation}
H=\diagram
        {\bf 1}  \ar@{-}[r] \ar@{-}[d]
&   \ar@{-}[r]  {\bf 4}    &  {\bf 5}     \\
        {\bf 2}  \ar@{-}[r]
& \ar@{-}[u] {\bf 3} \ar@{-}[r]       &  {\bf 6}\ar@{-}[u]   \\
\enddiagram
\quad\text{\rm and}\quad
H_0=\diagram
        {\bf 1}  \ar@{.}[r] \ar@{-}[d]
&   \ar@{-}[r]  {\bf 4}    &  {\bf 5}     \\
        {\bf 2}  \ar@{-}[r]
& \ar@{-}[u] {\bf 3} \ar@{.}[r]       &  {\bf 6}\ar@{-}[u]   \\
\enddiagram
\end{equation}
where the dotted lines represent missing edges. Clearly, $H$ is bipartite
and it is easy to check that indeed $H=H({\bf D},{\bf C},{\bf D},{\bf C},
{\bf D},{\bf C})$, $H$ holds. Therefore $H$ is a shape but $H_0$ is not.
Every sequence realizing $H_0$ has necessarily either
{\bf A} at {\bf 1}, and {\bf C} at {\bf 4} or vice versa. In the first case
{\bf D} is necessarily at {\bf 3} and {\bf 5}, which leaves no valid choice
for {\bf 6}. The second case follows analogously.

This is insofar remarkable since making the universal graph $H$
(being responsible for all interactions) more complex can simply
imply that not all of its subgraphs can be folded by sequences. This
is due, as the example indicates, to the nature of the base pairing
rule and shows clearly that both: $H$ and $\mathcal{R}$ determine what 
is a shape and what is not. 
For simple base graphs, like for instance $\mathcal{H}$, the lemma below 
shows that {\it any} subgraph (eq~(\ref{E:H})) is a shape. 
What we can deduce from this is (a) there exist many shapes and (b) 
$\mathcal{H}$ is so simple that it is indeed only $\mathcal{R}^\dagger$
that is relevant for the shapes. 
The result is

\begin{lemma}\label{L:bip}
Suppose $H$ is an arbitrary combinatorial graph over
$\{1,\dots,n\}$. \\
{\sf (a)} For any relation $\mathcal{R}$ any shape $\mathcal{S}$ is
bipartite.\\
{\sf (b)} For the relation $\mathcal{R}^\dagger$ and arbitrary base
graph $H$, any induced,
          bipartite subgraph of $H$ is a shape.\\
{\sf (c)} For the relation $\mathcal{R}^\dagger$ and the base graph
$\mathcal{H}$
          any $\mathcal{H}$-subgraph $H'$ is a shape.
\end{lemma}

\begin{figure}[ht]
\centerline{%
\epsfig{file=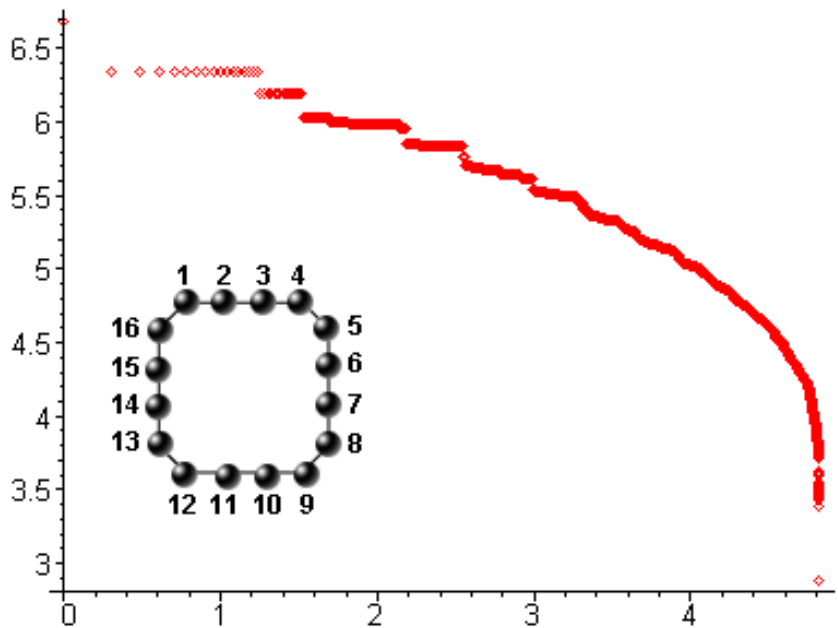,width=0.75\textwidth}\hskip20pt
}
\caption{\small A double logarithmic plot (base $10$) of the preimage
sizes of a combinatory map for $n=16$ as a function of the rank. The
underlying graph $\mathcal{H}_{16}$ is displayed in the lower right. The plot
shows that there are a few shapes with large and many shapes with very small
preimages. This observation is in complete analogy with RNA secondary structure
folding maps.
}
\label{fig:pre}
\end{figure}

Since any $\mathcal{H}$-subgraph is a shape we have for instance for sequences
of length $16$ exactly $2^{16}=65536$ different shapes in difference to only
$274$ RNA secondary structures realized by the minimum free energy folding
analyzed in \cite{Goebel:04}. This seems to indicate a vast difference between
combinatory maps and RNA secondary structure folding, however, closer 
inspection reveals that in fact most of these structures are very ``rare'', 
i.e.~only a few have large preimage sizes.
To understand what is happening we present in Figure~\ref{fig:pre} the data 
on the complete mapping from sequences of length $16$ into subgraphs of the 
cycle $\mathcal{H}_{16}$.
We plot the logarithm of the preimage sizes of a combinatory map over the
logarithm of the rank. We can deduce from Figure~\ref{fig:pre} that there 
are $393$ shapes with a preimage of size greater than $0.5\times 10^6$. The 
data on RNA secondary structures in \cite{Goebel:04} show that there are 
$132$ RNA minimum free energy structures with this property.
Figure~\ref{fig:pre} shows that combinatory maps exhibit $393$ shapes
with a preimage of size greater than $0.5\times 10^6$. As for RNA secondary
structures the data in \cite{Goebel:04} show that there are $132$ RNA minimum
free energy structures with this property. But what happens for larger 
sequence length? The asymptotics of RNA secondary 
structures \cite{Schuster:98,Reidys:07rna2} shows that the number of RNA 
secondary structures, ${\sf S}_2(n)$, satisfies ${\sf S}_2(n)\sim \kappa\, 
n^{-\frac{3}{2}}\alpha^n$ where $1.8488\le \alpha \le 2.64$, depending on 
what one considers a ``realistic'' secondary structure. In comparison a
combinatory map produces (Lemma~\ref{L:bip}) $2^n$ shapes.
Therefore combinatory maps produce a total number of structures
which is, for large $n$, in a comparable size-range.

The above observations motivate the question about the number of
shapes with large preimages \cite{Stroh:01}. For notational convenience let
\begin{equation}\label{number:1}
\mu_+=\left(\frac{1+\sqrt{5}}{2} \right)\qquad \text{\rm and}\qquad
\mu_-= \left(\frac{1-\sqrt{5}}{2} \right) \ .
\end{equation}
We next prove that there are many shapes with large preimages
\begin{lemma}\label{L:many}
Suppose the relation $\mathcal{R}^\dagger$ and the base graph
$\mathcal{H}$ are given, then there exist at least 
$\left(\sqrt{2}\right)^{n-1}$
shapes with the property that there are at least $2(\mu_+^n+\mu_-^n)$ sequences
folding into them.
\end{lemma}
Lemma~\ref{L:many} sets the stage for the further investigation of how this set
of sequences is organized. Now, knowing that there are exponentially large
sets of sequences realizing particular shapes what can be said about their
organization? Are they randomly distributed or clustered in sequence
space? What is their graph-structures as induced subgraphs of sequence space?


\section{Neutral networks of Combinatory Maps}


One difficulty in the context of neutral networks is that it is practically
impossible to prove they exist. Exhaustive enumeration of sequence spaces is
limited to small sequence length $n\le 20$ for four letter alphabets
\cite{Gruener:95a} and the results are of limited value since finite size
effects distort the picture. In case of
${\bf A},{\bf U},{\bf G},{\bf C}$-sequences about $60$\% of all sequences fold
into the open structure \cite{Goebel:04}. Several attempts have been made to
derive somewhat local criteria whether neutral networks exist \cite{Forst:00},
where the key idea is the probing for paths adopted from the actual random
graph proof in \cite{Reidys:97a,Reidys:02p2}.
In this context local parameters are the only quantities that give some clue 
about the existence and properties of neutral networks. 
In case of neutral networks modeled as random graphs, it is the number of 
neutral neighbors that controls global properties like connectivity
and density of the corresponding neutral network. A neutral neighbor is a
neighboring sequence which folds into the same structure and the fraction
\cite{Reidys:97p}
\begin{equation}
\lambda^* = 1-\sqrt[\alpha-1]{\alpha^{-1}}
\end{equation}
is actually the threshold value for connectivity and density. In the following
we can derive for combinatory maps the entire distribution of neutral neighbors
of particular shapes. The result is actually not ``local'' at all and entails
detailed information about the {\it entire} preimage of these shapes. To be
precise we can actually derive the underlying rational generating function
using the transfer matrix method of enumerative combinatorics.
We study the quantity $\lambda_{\mathcal{S}_M}(m)$ being the number of
sequences folding into the particular shape $\mathcal{S}$ having exactly $m$
neutral neighbors. Our result reads
\begin{theorem}\label{T:NN}
For arbitrary shape $\mathcal{S}_M$, where $M\subset \{1,\dots,k\}$ denotes its
set of isolated nucleotides, we have
\begin{equation}\label{E:ist}
\forall\, m\in \mathbb{N}\colon \lambda_{\mathcal{S}_M}(m) \ge
\lambda_{C_{2k}}(m)
\end{equation}
and the generating function of $\lambda_{C_{2k}}(m)$, $F(x,y)=\sum_{k\ge 2}
\sum_{m}\lambda_{C_{2k}}(m)x^{m}y^{2k}$ is given by
\begin{equation}\label{E:generate}
F(x,y)= \frac{2(-4x^{3}y^{6}+2x^{2}y^{6}+3x^{2}y^{4}-5+4x^{2}y^{2}+8xy^{2}-
6x^{3}y^{4}+2x^{4}y^{6})}{-2x^{3}y^{6}+x^{2}y^{6}+x^{2}y^{4}-1+2xy^{2}+
x^{2}y^{2}-2x^{3}y^{4}+x^{4}y^{6}}.
\end{equation}
\end{theorem}
The bi-variate function $F(x,y)$ provides detailed information about
neutral neighbors, of the entire preimages of shapes
$\mathcal{S}_M$. For instance, Taylor expansion of
eq.~(\ref{E:generate}) yields
$$
F(x,y)=10+(2x^2+4x)y^2+(12x^2+2x^4)y^4+(6x^2+16
x^3+12x^4+2x^6)y^6+O(y^{8})
$$
and the term $(12x^2+2x^4)y^4$ shows that for $n=4$ there are at least $12$
vertices with $2$ and $2$ vertices with $4$ neutral neighbors. Likewise, for
$n=6$, there are at least $6$ with $2$, $16$ with $3$, $12$ with $4$ and $2$
vertices with $6$ neutral neighbors. In addition eq.~(\ref{E:ist}) guarantees
that $\mathcal{H}$ itself provides a lower bound on the numbers of neutral
neighbors. I.e.~we can pinpoint a specific reference shape providing key 
information about the neutrality of the entire combinatory map.
\begin{figure}[ht]
\centerline{
\epsfig{file=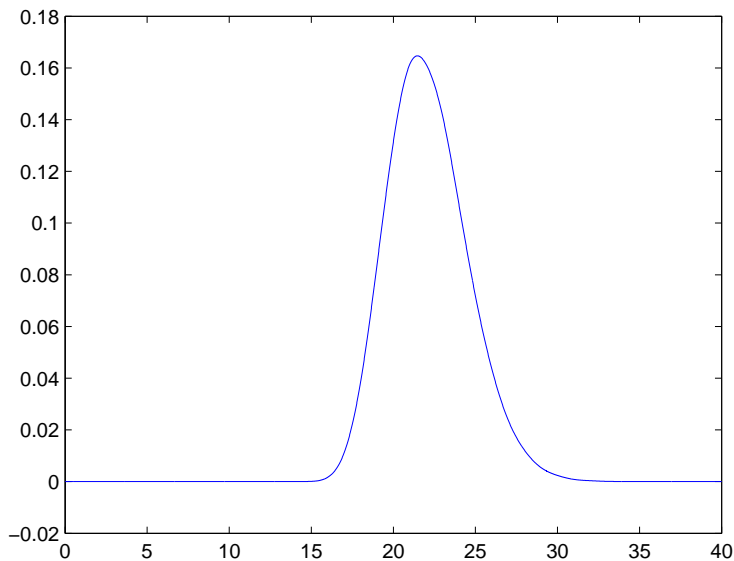,width=0.75\textwidth}\hskip20pt
}
\caption{\small The distribution of neutral neighbors for the entire
preimage of the ``reference'' shape $\mathcal{S}=\mathcal{H}_{40}$, 
where $n=40$ denotes the sequence length. 
We plot the fequency (y-axis) of numbers 
of neutral neighbors (x-axis) obtained from Theorem~\ref{T:NN}.
Note that the degree of a vertex in $Q_4^{40}$ is $120$, showing that
the lower bounds on the fractions of neutral neighbors range between 
$13$\% and $24$\% . }
\label{fig:$C_{40}$}
\end{figure}

In the previous section we have shown that there are many shapes
with large preimages. However, it is not obvious what the graph
structure of these preimages is. In this section we will study this
structure in detail and prove two remarkable properties. First there
are many shapes with sets of sequences having diameter $n$
i.e.~there exist two sequences which differ in {\it all} nucleotides
both of which map into the particular shape. This finding
is tantamount to percolation and indicates that the preimages are indeed
extended and not confined in some ``local'' region of sequence
space. Secondly we prove that the preimages of exponentially many shapes
contain large connected components. In other words we can
actually prove the existence of neutral networks for sequence to
shape maps, i.e.~many shapes have sets of sequences in which there exists 
a component of size $\ge \left(\sqrt{2}\right)^{n}$ and of diameter $n$.

\begin{figure}[ht]
\centerline{%
\epsfig{file=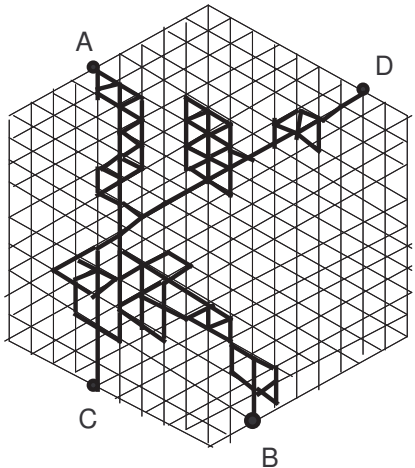,width=0.40\textwidth}\hskip15pt } \caption{\small
Neutral network. Sequence space is represented as lattice and the neutral net
is an induced subgraph (bold edges). We label the pairs of sequences 
representing antipodal pairs by $({\sf A},{\sf B})$ and $({\sf C},{\sf D})$.
The two key properties of neutral nets are their connectivity and 
percolation.} \label{F:2}
\end{figure}
\begin{theorem}\label{T:nn}{\bf (Neutral networks)}
Suppose the relation $\mathcal{R}^\dagger$ and the base graph $\mathcal{H}$ are
given. Then there exist at least $\left(\sqrt{2}\right)^{n-1}$ many shapes
$\mathcal{S}$ with the properties\\
{\sf (I)} the set of all sequences mapping into $\mathcal{S}$ has a connected
          component of size at least $\mu_+^n+\mu_-^n$.\\
{\sf (II)} the set of all sequences mapping into $\mathcal{S}$ percolates, i.e.
           has diameter $n$.\\
\end{theorem}
In comparison with the corresponding result for random graphs we
observe that the neutral networks are in fact slightly bigger and the
diameter indeed {\it equals} $n$. This is a result from the fact that the
simpler graph $\mathcal{H}$ allows for a different proof, which is very 
algorithmic. In fact the proof indicates how to
explicitly obtain these paths of diameter $n$, while the random
graph analogue can only produce their existence. In this sense both
constructions complement each other. To illustrate the idea of
Theorem~\ref{T:nn} we consider the cycle $\mathcal{H}_4$ and the
shape $\mathcal{S}=\mathcal{H}_4$. Then we have the following situation (using
the notation of the proof of Theorem~\ref{T:nn})
$$
a^\varnothing=({\bf C},{\bf D},{\bf C},{\bf D}) \quad \text{\rm and }
\quad C_4(({\bf C},{\bf D},{\bf C},{\bf D}))=C_4 \ .
$$
Theorem~\ref{T:nn} guarantees the existence of the antipodal sequence
$\tilde{a}^\varnothing=({\bf B},{\bf A},{\bf B},{\bf A})$ and a path
connecting $a^\varnothing$ and $\tilde{a}^\varnothing$ obtained via
the steps {\sf (a)}, {\sf (b)} and {\sf (c)}. Explicitly this path
for $\mathcal{S}_\varnothing$ from $a^\varnothing$ to
$\tilde{a}^\varnothing$ is given by {\small
$$
\underbrace{
\diagram
\fbox{{\bf C}}  \ar@{-}[r]      & {\bf D} \\
      {\bf D}  \uline\rline   & {\bf C}\uline \\
\enddiagram
\ \mapsto
\diagram
      {\bf B}  \ar@{-}[r]      & {\bf D} \\
      {\bf D}  \uline\rline    & \fbox{{\bf C}}\uline \\
\enddiagram}_{\text{\rm step}
 {\sf (a)}: \,\text{\rm replace {\bf C} by {\bf B}}}
\ \mapsto
\underbrace{\diagram
      {\bf B}  \ar@{-}[r]      & {\bf D} \\
      \fbox{{\bf D}}  \uline\rline    & {\bf B}\uline\\
\enddiagram
\ \mapsto
\diagram
      {\bf B}  \ar@{-}[r]      & \fbox{{\bf D}} \\
      {\bf A}  \uline\rline    & {\bf B}\uline\\
\enddiagram}_{{\rm step} {\sf (b)}: \,\text{\rm replace {\bf D} by {\bf A}}}
\ \mapsto
\diagram
      {\bf B}  \ar@{-}[r]      & {\bf A} \\
      {\bf A}  \uline\rline    & {\bf B}\uline\\
\enddiagram
$$
}

Theorem \ref{T:nn} holds for many shapes. For instance the neutral path for
$\mathcal{S}_{\{1\}}$, which has length ${\sf diam}(Q_4^4)=4$ and
which connects the sequences $a^{\{1\}},\tilde{a}^{\{1\}}$ is given
by {\small
$$
\underbrace{
\diagram
      {\bf A}  \ar@{.}[r]         & {\bf D} \\
      {\bf D}  \ar@{.}[u]\rline   &\fbox{{\bf C}}\uline \\
\enddiagram
}_{{\rm step} {\sf (a)}: \text{\rm replace {\bf C} by {\bf B}}}
\ \mapsto
\underbrace{\diagram
      {\bf A}  \ar@{.}[r]      &     \fbox{{\bf D}} \\
      {\bf D}  \ar@{.}[u]\rline    & {\bf B}\uline \\
\enddiagram
\ \mapsto
\diagram
      {\bf A}  \ar@{.}[r]      &       {\bf A} \\
\fbox{{\bf D}}  \ar@{.}[u]\rline    &   {\bf B}\uline\\
\enddiagram}_{{\rm step} {\sf (b)}: \,\text{\rm replace {\bf D} by {\bf A}}}
\ \mapsto
\underbrace{
\diagram
     \fbox{ {\bf A}}  \ar@{.}[r]      & {\bf A} \\
      {\bf A}  \ar@{.}[u]\rline    & {\bf B}\uline\\
\enddiagram}_{{\rm step} {\sf (c)}: \,\text{\rm replace {\bf A} by {\bf C}}}
\ \mapsto
\diagram
      {\bf C}  \ar@{.}[r]      & {\bf A} \\
      {\bf A}  \ar@{.}[u]\rline    & {\bf B}\uline\\
\enddiagram
$$
}


\section{Appendix}

{\bf Proof of Lemma}~\ref{L:bip}
To show {\sf (a)} we first prove that for any relation satisfying
eq.~(\ref{E:1}), eq.~(\ref{E:2})
and eq.~(\ref{E:3}) a shape $\mathcal{S}$ is bipartite. \\
{\it Claim.} Any closed walk in $\mathcal{S}$ has even length.\\
Since $\mathcal{S}$ is a shape we have $\mathcal{S}=H(v)$, whence
for any closed walk $w=(w_1,w_2\dots,w_r,w_1)$ in $\mathcal{S}$
there exists at least one sequence
$x=(x_{w_1},x_{w_2},\dots,x_{w_r},x_{w_1})$, where $x_h\in
\{\bf{A},\bf{U},\bf{G},\bf{C}\}$. Therefore there exists an
injection
\begin{eqnarray*}
\{(x_{w_1},x_{w_2},\dots,x_{w_r},x_{w_1})\mid w \
\text{\rm is a closed walk in $\mathcal{S}$}\}
\longrightarrow \{\gamma \mid \text{\rm $\gamma$ is a closed walk in
$G(\mathcal{R})$}\}
\end{eqnarray*}
The idea is to show that
$$
\{\gamma \mid \text{\rm $\gamma$ is a closed walk in $G(\mathcal{R})$ of odd
length}\}=\varnothing   \  .
$$
Suppose $\gamma$ is a closed walk of minimal, odd length in
$G(\mathcal{R})$. Obviously, there are only $4$ vertices in
$G(\mathcal{R})$. We can conclude from this that $\gamma$
contains a cycle of length $3$ which is in view of eq.~(\ref{E:3})
impossible, whence the claim.\\
We next select an arbitrary vertex, $i\in \{1\,\dots n\}$ and color
all vertices in even distance to $i$ blue and all vertices in odd distance
red. Suppose this procedure leads to two monochromatic adjacent
vertices $j,r$. Then we obtain a closed walk containing $i,j$ and
$r$ of odd length. By induction we can conclude that this walk
contains a cycle of odd length, which is impossible, whence
$\mathcal{S}$ is bipartite and
assertion {\sf (a)} follows.\\
Next we show {\sf (b)} by constructing a vertex
$v=(x_1,\dots,x_n)\in Q_4^n$ with the property
$H_{\mathcal{R}_{NC}}(v)=H'$, where $H'$ is an arbitrary induced,
bipartite subgraph of $H$. Since $H'$ is induced in $H$ there exists
some set $M\subset \{1,\dots,n\}$ such that $E_{H'}=\{\{i,j\}\mid
\{i,j\}\in E_H\,\wedge\, i,j\in M\}$. First, for all coordinates
$x_j$ where $j\not \in M$ we set $x_j={\bf A}$. Then by definition
of $\mathcal{R}^\dagger$ for $i,i'\not \in M$, $\{x_i,x_{i'}\}\not\in
\mathcal{R}^\dagger$ holds. Since $H'$ is bipartite there exists for
the vertices $j\in M$ a bi-coloring (red/blue) such that no two
$H'$-adjacent vertices are monochromatic. Suppose $x_j,x_k$ are
coordinates where $j,k\in M$. We choose a bi-coloring (red/blue) and
set $x_j={\bf D}$ for $j$ being colored red and $x_k={\bf C}$ for
$k$ being colored blue, respectively. In view of $({\bf D},{\bf
C}),({\bf C},{\bf D})\in \mathcal{R}^\dagger$, we can conclude that for
$j,k\in M$ and $\{j,k\}\in H$ we have $\{x_j,x_{k}\}\in
\mathcal{R}^\dagger$. Since $({\bf A},{\bf C}),({\bf A},{\bf D})\not
\in \mathcal{R}^\dagger$ we derive that for $i\not\in M$ and $j\in M$,
$\{x_i,x_j\}\not \in\mathcal{R}^\dagger$ holds. Therefore
$H_{\mathcal{R}^\dagger}((x_1,\dots,x_n))=H'$ i.e.~any
induced bipartite subgraph of $H$ is a shape.\\
Next we show {\sf (c)}, i.e.~for $\mathcal{H}$ (eq~(\ref{E:H})) any
$H'<\mathcal{H}$ is a shape. We proceed by explicitly constructing a
vertex $v=(x_1,\dots,x_n)\in Q_4^n$ with the property
$\mathcal{H}_{\mathcal{R}^\dagger}(v)=H'$. W.l.o.g.~we can assume
that $n$ is even since the isolated point $u$ does not contribute to
the $\mathcal{H}$-shapes. Then we have $\mathcal{H}=C_{2k}$ and
$V_{C_{2k}}=\{1,\dots,2k\}$. We label the $H'$-vertices $\{1,\dots,
2k\}$ clock-wise such that the (clockwise) first vertex in one
largest $H'$-component is $1$. Then $H'$ corresponds to a unique
sequence of components. We assume now $x_i\in\{{\bf A},{\bf B}\}$
and label all $H'$-vertices except of those contained in the
component proceeding vertex $1$. We set inductively
\begin{equation}
x_i=
\begin{cases}
{\bf A}            & \text{\rm iff $i=1$}\\
x_{i-1}            & \text{\rm iff $\{i-1,i\}$ is not an edge in $H'$}\\
\overline{x_{i-1}} & \text{\rm iff $\{i-1,i\}$ is an edge in $H'$} \ ,
\end{cases}
\end{equation}
where $\overline{{\bf B}}={\bf A}$ and $\overline{{\bf A}}={\bf B}$.
As for the labeling of the component preceding the component containing
vertex $1$, we start with $x_j={\bf C}$ and continue inductively
$x_{j+1}={\bf D},x_{j+2}={\bf C},\dots $. This procedure
results in a
labeling compatible with $H'$ since for $\{i-1,i\}\in H'$ we have either
$\{{\bf C},{\bf D}\}$ or $\{{\bf A},{\bf B}\}$ and for $\{i-1,i\}\not\in H'$
we have $\{{\bf A},{\bf A}\}$, $\{{\bf B},{\bf B}\}$ and $\{{\bf A},{\bf C}\}$
or $\{{\bf B},{\bf C}\}$ (at the beginning of the last component) and
$\{{\bf D},{\bf A}\}$ or $\{{\bf C},{\bf A}\}$ (at the end of the
last component). Accordingly we obtain a sequence $\tilde{v}_{H'}$
with the property $\mathcal{H}(\tilde{v}_{H'})=H'$.$\ \square$


{\bf Proof of Lemma}~\ref{L:many}
By definition, there exists a unique component of $\mathcal{H}$
which is a cycle of even length, $C_{2k}$. $C_{2k}$ contains for $n$
even all and for $n$ odd all but one $\mathcal{H}$-vertices. Suppose
$C_{2k}$ contains the vertices
$\{i_1,j_1,\dots,i_{k},j_k\}$, where $i_1<j_1<i_2<\dots i_{k}<j_k$. \\
{\it Claim.} The number of $2k$-tuples $(x_{i_1},x_{j_1},\dots,x_{i_{k}},
x_{j_k})$ such that $C_{2k}((x_{i_1},x_{j_1}\dots,x_{i_{k}},x_{j_k}))=
C_{2k}$ i.e.
$
(x_{i_1},x_{j_1},\dots,x_{i_{k}},x_{j_k})\in\vartheta_{C_{2k}}^{-1}(C_{2k})
$
is given by
\begin{equation}\label{E:precise}
2\, \left(\mu_+^{2k}+\mu_-^{2k}\right) \ .
\end{equation}
To prove the claim we observe that $\mathcal{R}^\dagger$ induces the
digraph $D_{\mathcal{R}^\dagger}$ defined as follows: {\small
$$
D_{\mathcal{R}^\dagger}= \diagram
{\bf A}  \ar@/^1pc/@{<-}[rr]|{} \ar@/_1pc/@{->}[rr]|{}& & {\bf B} \\
{\bf C} \ar@/^1pc/@{<-}[rr]|{} \ar@/_1pc/@{->}[rr]|{}& & {\bf D}
\ar@/^1pc/@{<-}[u]|{} \ar@/_1pc/@{->}[u]|{}
\enddiagram  \qquad\text{\rm and} \quad  A_{D_{\mathcal{R}^\dagger}}=
\bordermatrix{%
  & {\bf A} & {\bf B} & {\bf D} & {\bf C} \cr%
  & 0 & 1 & 0 & 0  \cr%
  & 1 & 0 & 1 & 0  \cr%
  & 0 & 1 & 0 & 1  \cr%
  & 0 & 0 & 1 & 0 \cr%
}%
$$}
The number of $2k$-tuples $(x_{i_1},x_{j_1}\dots,x_{i_{k}},x_{j_k})$
with the property
$C_{2k}((x_{i_1},x_{j_1},\dots,x_{i_{k}},x_{j_k}))=C_{2k}$ is equal
to the number of closed walks of length $2k$ in
$D_{\mathcal{R}^\dagger}$. Indeed, in order to obtain such an
$2k$-tuple we fix an index, $i_1$, say. Then we start with
successively ${\bf A}$, ${\bf B}$, ${\bf D}$ and ${\bf C}$ and form
of closed walks of length $2k$ in $D_{\mathcal{R}^\dagger}$ starting
and ending at ${\bf A}$, ${\bf B}$, ${\bf D}$ and ${\bf C}$. All
these walks are counted respectively, since we have labeled graphs.
The number of closed walks of length $\ell$ in
$D_{{\mathcal{R}}_{NC}}$ starting and ending at $i$ is given by
$(A_{D_{{\mathcal{R}}_{NC}}}^\ell)_{i,i}$, whence the number of all
closed walks of length $\ell$ is simply ${\sf
Tr}(A_{D_{{\mathcal{R}^\dagger}}}^\ell) =\sum_i
(A_{D_{{\mathcal{R}^\dagger}}}^\ell)_{i,i}$. From the definition of the
characteristic polynomial,~i.e.~${\sf
Tr}(A_{D_{{\mathcal{R}^\dagger}}}^\ell)=\omega_1^\ell+\dots
+\omega_r^\ell$, where $\omega_1,\dots,\omega_r$ are the
eigenvalues of $A_{D_{{\mathcal{R}^\dagger}}}$ (note $r=4$). We obtain
\begin{eqnarray*}
\sum_{\ell\ge 0}{\sf Tr}(A_{D_{{\mathcal{R}}_{NC}}}^\ell) z^\ell & =
&
\sum_{\ell\ge 0} \left[\omega_1^\ell+\dots +\omega_r^\ell \right]z^\ell \\
 & = & \sum_{\ell\ge 0}
\left[(1+(-1)^\ell)\left(\mu_+^\ell+
                     \mu_-^\ell\right)\right] z^\ell
\end{eqnarray*}
and the claim follows.\\
Suppose $(x_{i_1},x_{j_1},\dots,x_{i_{k}},x_{j_k})\in \vartheta_{C_{2k}}^{-1}
(C_{2k})$ and $M\subset \{1,\dots,k\}$. We consider the involution
$\tau\colon \mathcal{A}\rightarrow  \mathcal{A}$, where $\tau({\bf A})=
{\bf B}$ and $\tau({\bf D})={\bf C}$ and set
\begin{eqnarray}\label{E:I_M}
\quad I_M(x_{i_1},x_{j_1}\dots,x_{i_{k}},x_{j_k}) & = &
(y_{i_1},x_{j_1}\dots,y_{i_{k}},x_{j_k}), \ \text{\rm where} \
y_{i_\ell}=
\begin{cases}
\tau(x_{i_\ell}) & \text{\rm for } i_\ell \in M         \\
x_{i_\ell}       & \text{\rm for } i_\ell \not\in M \ .
\end{cases}
\end{eqnarray}
{\it Claim.} There exists a bijection
$$
\beta\colon \{M\subset \{1,2,\dots,k\}\}\rightarrow \{\mathcal{S}_M\}, \quad
M\mapsto\mathcal{S}_{M}
$$
where $\mathcal{S}_M$ is obtained by deleting any two $C_{2k}$-edges
incident to the vertices $i_{h}\in M$ and
\begin{equation}
\forall \,(x_{i_1},x_{j_1}\dots,x_{i_{k}},x_{j_k})
\in \vartheta_{C_{2k}}^{-1}(C_{2k});\,\quad
\mathcal{S}_M=C_{2k}(I_M(x_{i_1},x_{j_1}\dots,x_{i_{k}},x_{j_k})) \ .
\end{equation}
Suppose $M\neq M'$ then w.l.o.g.~we can assume that there exists
some index $i_{h}\in M\setminus M'$, i.e.~$i_{h}$ is isolated in
$\mathcal{S}_{M}$ but not in $\mathcal{S}_{M'}$. Since $j_{h-1}$ and
$j_{h}$ are both in $\mathcal{S}_M$ and $\mathcal{S}_{M'}$ we have
$\{j_{h-1},i_h\},\{j_{h},i_h\},\in \mathcal{S}_{M'}$ but not in
$\mathcal{S}_{M}$, whence $\mathcal{S}_{M}$ and $\mathcal{S}_{M'}$
are different shapes. Since $\mathcal{S}_M$ is an induced bipartite
subgraph, Lemma~\ref{L:bip} implies that any $\mathcal{S}_M$ is a
shape. When $i_{h}\in M$ the following diagram {\small
$$
\diagram
               &         &  x_{j_h} \\
x_{j_{h-1}} \ar@{-}[r] & \text{\rm \fbox{$x_{i_{h}}$}}
\ar@{-}[ur] \ar@{-}[r] & x_{j_{h}} \\
\enddiagram
\quad \mapsto\quad
\diagram
                         &                                           & {x_{j_{h}}} \\
{x_{j_{h-1}}} \ar@{.}[r] & \text{\rm \fbox{$\tau(x_{i_{h}})$}}
\ar@{.}[ur] \ar@{.}[r] & {x_{j_{h}}}
\enddiagram
$$}
shows that $I_M$ has the property: for arbitrary
$$
(x_{i_1},x_{j_1}\dots,x_{i_{k}},x_{j_k})\in\vartheta_{C_{2k}}^{-1}(C_{2k})
$$
the shape $C_{2k}(I_M(x_{i_1},x_{j_1}\dots,x_{i_{k}},x_{j_k}))$
differs from $C_{2k}$ exactly by deleting the two $C_{2k}$-edges
incident to all $i_\ell\in M$; explicitly {\tiny
$$
\diagram
               &         &  {\bf D} \\
{\bf A} \ar@{-}[r] & \fbox{{\bf B}} \ar@{-}[ur] \ar@{-}[r] & {\bf A} \\
\enddiagram
\ \mapsto \
\diagram
               &         &  {{\bf D}} \\
{\bf A} \ar@{.}[r] & \fbox{{\bf A}} \ar@{.}[ur] \ar@{.}[r] & {\bf A} \\
\enddiagram
\quad
\diagram
               &         &  {\bf B} \\
{\bf C} \ar@{-}[r] & \fbox{{\bf D}} \ar@{-}[ur] \ar@{-}[r] & {\bf C}\\
\enddiagram
\ \mapsto \
\diagram
               &         &  {\bf B} \\
{\bf C} \ar@{.}[r] & \fbox{{\bf C}} \ar@{.}[ur] \ar@{.}[r] & {\bf C}\\
\enddiagram
\quad
$$
}
{\tiny
$$
\diagram
                                 &  \fbox{{\bf A}}\ar@{-}[r]  &  {\bf B}  \\
{\bf B}\ar@{-}[ur] \ar@{-}[r]  & \fbox{{\bf D}} \ar@{-}[ur]
\ar@{-}[r] &
{\bf C}\\
\enddiagram
\ \mapsto \
 \diagram
                                 & \fbox{{\bf B}}\ar@{.}[r]  &  {\bf B}  \\
{\bf B}\ar@{.}[ur] \ar@{.}[r]&\fbox{{\bf C}} \ar@{.}[ur] \ar@{.}[r] & {\bf C}\\
\enddiagram
\quad
\diagram
                                 &  \fbox{{\bf C}}\ar@{-}[r]  &  {\bf D}  \\
{\bf D}\ar@{-}[ur] \ar@{-}[r]&\fbox{{\bf B}} \ar@{-}[ur] \ar@{-}[r] & {\bf A}\\
\enddiagram
\ \mapsto \
 \diagram
                               &  \fbox{{\bf D}}\ar@{.}[r]  &  {\bf D}  \\
{\bf D}\ar@{.}[ur] \ar@{.}[r]  &\fbox{{\bf A}} \ar@{.}[ur]
\ar@{.}[r] & {\bf A}\\
\enddiagram
$$
}
and the claim is proved. The claim implies that $I_M$ induces the injection
\begin{eqnarray}
\quad I_M \colon \vartheta_{C_{2k}}^{-1}(C_{2k}) & \longrightarrow &
\vartheta_{C_{2k}}^{-1}(\mathcal{S}_M), \ \quad
(x_{i_1},x_{j_1}\dots,x_{i_{k}},x_{j_k})\mapsto
I_M(x_{i_1},x_{j_1}\dots,x_{i_{k}},x_{j_k}) \ .
\end{eqnarray}
This injection allows us to relate the sets $\vartheta_{C_{2k}}^{-1}(C_{2k})$
and $\vartheta_{C_{2k}}^{-1}(\mathcal{S}_M)$ and in particular
\begin{equation}
\vert\vartheta_{C_{2k}}^{-1}(C_{2k})\vert
\le \vert\vartheta_{C_{2k}}^{-1}(\mathcal{S}_M)\vert \ .
\end{equation}
Since $M\subset \{1,\dots,k\}$ was arbitrary we can conclude that there
are $2^k$ subsets and hence $2^k$ distinct shapes $\mathcal{S}_M$.
Hence there exist at least
$$
2^{k} \ge  \left(\sqrt{2}\right)^{n-1}
$$
shapes $\mathcal{S}$ with the property
\begin{eqnarray*}
\vert \vartheta_\mathcal{H}^{-1}(\mathcal{S})\vert  \ge
                                \vert \vartheta_\mathcal{H}^{-1}(\mathcal{H})\vert
 \ge  2\, \left(\mu_+^{2k}+
                   \mu_-^{2k}\right) \ .
\end{eqnarray*}
In case of $n\not\equiv 0\mod 2$ we have exactly one more isolated point,
i.e.
\begin{equation}\label{E:isolated}
\vert \vartheta_\mathcal{H}^{-1}(\mathcal{S})\vert\ge 8\,
\left(\mu_+^{n-1}+
                   \mu_-^{n-1}\right)
\end{equation}
and since $4\ge \left( \mu_+ + \mu_- \right)$
the lemma follows. $\ \square$


{\bf Proof of Theorem}~\ref{T:nn}
We first prove that at least $\left(\sqrt{2}\right)^{n-1}$ shapes $\mathcal{S}$
have a preimage $\vartheta_\mathcal{H}^{-1}(\mathcal{S})$ with diameter $n$.
We will work with the particular set of shapes
$\{\mathcal{S}_M \mid M\subset \{1,\dots,k\}\}$, introduced in
Lemma~\ref{L:many} and prove that all of them have a
component of size $\ge \mu_+^{n}+ \mu_-^{n} >\left(\sqrt{2}\right)^{n}$ and
${\sf diam}(\vartheta_\mathcal{H}^{-1}(\mathcal{S}))=n$.
Let $C_{2k}$ be the $\mathcal{H}$-cycle, which contains all
$\mathcal{H}$-vertices for $n$ even and all but one $\mathcal{H}$-vertices,
for $n$ odd.
Let $V_{C_{2k}}=\{i_1,j_1,\dots,i_{k},j_k\}$, where
$i_1<j_1<i_2<\dots i_{k}<j_k$. \\
{\it Claim $1$.} Let $M\subset \{1,\dots,k\}$, then there exist at
least $2^{k}$ shapes $S_M$ over $Q_4^{2k}$ such that
\begin{equation}
\text{\sf diam}(\vartheta_\mathcal{H}^{-1}(\mathcal{S}_M))=
\begin{cases}
n   & \text{\rm for}\  n\equiv 0\mod 2    \\
n-1 & \text{\rm for}\  n\not\equiv 0\mod 2 \ .
\end{cases}
\end{equation}
We first show that for each $M$ there exists a pair of antipodal sequences,
i.e.~$(a^M,\tilde{a}^M)$ with $d(a^M,\tilde{a}^M)=2k$ and a path
$(a^M,w_1^M,\dots,w_{2k-1}^M,\tilde{a}^M)$ such that
$\vartheta_{C_{2k}}(w_i^M)=\mathcal{S}_M$.
\begin{equation}\label{E:a^M}
a^M=(a^M_{i_1},a_{j_1}\dots,a^M_{i_{k}},a_{j_k}), \quad \text{\rm where}\quad
a_{j_h}={\bf D}, \ \text{\rm and}\  a^M_{i_h}=
\begin{cases}
{\bf A} & \text{\rm for} \ i_h\in M \\
{\bf C} & \text{\rm otherwise.}
\end{cases}
\end{equation}
In particular we have $a^\varnothing=({\bf C},{\bf D},\dots, {\bf
C},{\bf D})$. Then $\mathcal{S}_M=C_{2k}(a^M)$, i.e.~$\mathcal{S}_M$
is the shape obtained by removing for each $i_h\in M$ the two
incident $C_{2k}$-edges. Next we define an antipode $\tilde{a}^M$,
i.e.~an element of $Q_4^{2k}$ with the property
$d(a^M,\tilde{a}^M)=2k$ as follows
\begin{equation}
\tilde{a}^M=(\tilde{a}^M_{i_1},\tilde{a}_{j_1}\dots,\tilde{a}^M_{i_{k}},
\tilde{a}_{j_k}), \quad \text{\rm where}\quad
\tilde{a}_{j_h}={\bf A}, \ \text{\rm and}\ \tilde{a}^M_{i_h}=
\begin{cases}
{\bf C} & \text{\rm for} \ i_h\in M \\
{\bf B} & \text{\rm otherwise.}
\end{cases}
\end{equation}
We can transform $a^M$ into $\tilde{a}^M$ by successively changing
exactly one coordinate in three steps: {\sf (a)} replace (in any
order) for $i_h\not\in M$ successively all $a_{i_h}={\bf C}$ by
${\bf B}$, {\sf (b)} replace (in any order) successively all
$a_{j_h}={\bf D}$ by ${\bf A}$ and finally {\sf (c)} substitute (in
any order) for all $i_h\in M$ $a_{i_h}={\bf A}$ by
${\bf C}$. \\
This proves that there exists a $Q_4^{2k}$-path
\begin{equation}
(a^M,w_1^M,\dots,w_{2k-1}^M,\tilde{a}^M)
\end{equation}
connecting $a^M$ and $\tilde{a}^M$, such that
\begin{equation}
\forall \, 1\le i\le 2k-1, \quad C_{2k}(w_i^M)=\mathcal{S}_M \ .
\end{equation}
I.e.~all intermediate steps of the path are mapped by
$\vartheta_{\mathcal{H}}$ into
the shape $\mathcal{S}_M$.
As shown in Lemma~\ref{L:many} there are $2^k$ different shapes
$\mathcal{S}_M$
induced by the subsets $M\subset \{1,\dots,k\}$, whence Claim $1$.\\
In case of $n\equiv 0\mod 2$ we derive
$2^k=\left(\sqrt{2}\right)^n$. In case of $n\not \equiv 0\mod 2$
there exists exactly one vertex $u$ which is isolated in
$\mathcal{H}$. Then we simply add the isolated point $u$ to each
shape $\mathcal{S}_M$ and shall in the following identify these new
shapes with $\mathcal{S}_M$. Then
$\vert\vartheta_\mathcal{H}^{-1}(\mathcal{S}_M)\vert =
4\vert\vartheta_{C_{2k}}^{-1}(\mathcal{S}_M)\vert$. We can choose
$a_u={\bf A}$ and $\tilde{a}_u={\bf B}$ and
\begin{eqnarray*}
a_u^M & = & (a^M_{i_1},a_{j_1}\dots,a_u,\dots, a^M_{i_{k}},a_{j_k})\\
\tilde{a}_u^M & = & (\tilde{a}^M_{i_1},\tilde{a}_{j_1}\dots,\tilde{a}_u,\dots,
\tilde{a}^M_{i_{k}},\tilde{a}_{j_{k}})
\end{eqnarray*}
satisfy $d(a_u^M,\tilde{a}_u^M)=n$ and there exists a $Q_4^{n}$-path
$(a_u^M,w_1^M,\dots,w_{2k}^M,\tilde{a}_u^M)$ connecting $a_u^M$ and
$\tilde{a}_u^M$, with the property
\begin{equation}
\forall \, 1\le i\le 2k, \quad C_{2k}(w_i^M)=\mathcal{S}_M \ .
\end{equation}
Therefore we have proved that at least $\left(\sqrt{2}\right)^{n-1}$
shapes $\mathcal{S}_M$ have a preimage $\vartheta_\mathcal{H}^{-1}(\mathcal{S}_M)$
with diameter $n$.\\
{\it Claim $2$.}
\begin{equation}\label{E:claim2}
\vert \left\{\mathcal{S}_M\mid \vert \mathcal{C}(\vartheta_\mathcal{H}^{-1}
(\mathcal{S}))\vert \ \ge
\mu_+^{2k}+\mu_-^{2k}
\right\}\vert \ge 2^k \ .
\end{equation}
To prove the Claim $2$ we first observe that
$\vartheta_\mathcal{H}^{-1}(\mathcal{H})$ has
exactly two components of equal size
\begin{equation}\label{E:cool2}
\mu_+^{2k}+\mu_-^{2k} \ .
\end{equation}
Indeed, any vertex $v\in \vartheta_\mathcal{H}^{-1}(\mathcal{H})$ can be
transformed into either
$$
a^\varnothing=({\bf C},{\bf D},{\bf C},\dots,{\bf D},{\bf C}),
\quad \text{\rm or}
\quad
b^\varnothing=({\bf D},{\bf C},\dots,{\bf D},{\bf C},{\bf D})
$$
successively using
the two steps {\sf (I)} replace (in any order) all ${\bf A}$ by ${\bf D}$
and {\sf (II)} replace all (in any order) ${\bf B}$ by ${\bf C}$. Hence
there exist exactly two components and the map
$$
\sigma(x_{i_1},x_{j_1},\dots, x_{i_k},x_{j_k})=
(x_{j_k},x_{i_1},\dots, x_{j_{k-1}},x_{i_k})
$$
is a bijection between them, whence they have equal size.
Eq.~(\ref{E:cool2}) then follows from eq.~(\ref{E:precise}) in
Lemma~\ref{L:many}. We next claim that the mapping
$I_M$ of eq.~(\ref{E:I_M}) is in fact an injective graph morphism
\begin{eqnarray}
\ I_M\colon \vartheta_{C_{2k}}^{-1}(C_{2k}) & \longrightarrow &
\vartheta_{C_{2k}}^{-1}(\mathcal{S}_M), \ \quad
(x_{i_1},x_{j_1}\dots,x_{i_{k}},x_{j_k})\mapsto
I_M(x_{i_1},x_{j_1}\dots,x_{i_{k}},x_{j_k}).
\end{eqnarray}
I.e. for two adjacent vertices $v,v'\in\vartheta_{C_{2k}}^{-1}$, the
vertices $I_{M}(v)$ and $I_{M}(v')$ are adjacent. To prove
this we consider the diagrams {\small
$$
x_{j_{h-1}}=x_{j_h}={\bf B}:\quad
\underbrace{\diagram
                                 &  \fbox{{\bf A}}\ar@{-}[r]  &  {\bf B}  \\
{\bf B}\ar@{-}[ur] \ar@{-}[r]  & \fbox{{\bf D}}
\ar@/^2pc/@{->}[u]|{}
\ar@/_2pc/@{<-}[u]|{}
\ar@{-}[ur]  & \\
\enddiagram}_{(x_{j_{h-1}},x_{i_h},x_{j_{h}})}
\qquad \mapsto
\qquad
\underbrace{\diagram
                                 &  \fbox{{\bf B}}\ar@{.}[r]  &  {\bf B}  \\
{\bf B}\ar@{.}[ur] \ar@{.}[r]  & \fbox{{\bf C}} \ar@{.}[ur]
\ar@/^2pc/@{->}[u]|{}
\ar@/_2pc/@{<-}[u]|{}
  & \\
\enddiagram}_{(x_{j_{h-1}},x_{i_h},x_{j_{h}})}
$$}
{\small
$$
x_{j_{h-1}}=x_{j_h}={\bf D}:\quad
\underbrace{\diagram
                                 &  \fbox{{\bf C}}\ar@{-}[r]  &  {\bf D}  \\
{\bf D}\ar@{-}[ur] \ar@{-}[r]&\fbox{{\bf B}} \ar@{-}[ur]
\ar@/^2pc/@{->}[u]|{}
\ar@/_2pc/@{<-}[u]|{}
&\\
\enddiagram}_{(x_{j_{h-1}},x_{i_h},x_{j_{h}})}
\qquad \mapsto
\qquad
\underbrace{\diagram
                            &  \fbox{{\bf D}}\ar@{.}[r]  &  {\bf D}  \\
{\bf D}\ar@{.}[ur] \ar@{.}[r]&\fbox{{\bf A}} \ar@{.}[ur]
\ar@/^2pc/@{->}[u]|{}
\ar@/_2pc/@{<-}[u]|{}
&\\
\enddiagram}_{(x_{j_{h-1}},x_{i_h},x_{j_{h}})}
$$}
The above diagrams represent the two scenarios for two adjacent
vertices $v,v'\in\vartheta^{-1}_{C_{2k}}(C_{2k})$. I.e.~if $v$ and
$v'$ are both contained in $\vartheta_{C_{2k}}^{-1}(C_{2k})$ and
differ in $x_{i_h}$ and $x_{i_h}'$ then we have either
$x_{j_{h-1}}=x_{j_h}={\bf B}$ and $x_{i_h}={\bf D}$ and
$x_{i_h}'={\bf A}$ or $x_{j_{h-1}}=x_{j_h}={\bf D}$ and
$x_{i_h}={\bf B}$ and $x_{i_h}'={\bf C}$. Suppose we apply $I_{M}$
and ${i_h}\in M$, then the resulting vertices $I_{M}(v)$ and
$I_{M}(v')$ are again adjacent, whence $I_M$ is an injective graph
morphism. \ Accordingly, $I_M$ maps components into components, from
which we can conclude that for each $M\subset \{1,\dots,k\}$ the
shape $\mathcal{S}_M$ has a component of size
$\mu_+^{2k}+\mu_-^{2k}$ and Claim $2$ is proved.\\
In case of $2k=n$ the
assertion follows directly. For $n$ odd we have to repeat the
argument in Lemma~\ref{L:many}, where we considered the isolated
point $u$ in eq.~(\ref{E:isolated}). Since we used the same set of
shapes $\{\mathcal{S}_M\mid M\subset \{1,\dots,k\}\}$ for both
claims the theorem follows. $\ \square$

{\bf Proof of Theorem}~\ref{T:NN}
It is clear that we can restrict our analysis to the case $n\equiv
0\mod 2$, i.e.~$\mathcal{H}=C_{2k}$, since the isolated point
contributes always $4$ neutral neighbors for any shape.
Eq.~(\ref{E:ist}) is a direct consequence of
\begin{eqnarray*}
\ I_M\colon \vartheta_{C_{2k}}^{-1}(C_{2k}) & \longrightarrow &
\vartheta_{C_{2k}}^{-1}(\mathcal{S}_M), \ \quad
(x_{i_1},x_{j_1}\dots,x_{i_{k}},x_{j_k})\mapsto
I_M(x_{i_1},x_{j_1}\dots,x_{i_{k}},x_{j_k}) \ .
\end{eqnarray*}
being an injective graph morphism. Thus it suffices to prove
eq.~(\ref{E:generate}). We observe that for $v\in
\vartheta_{C_{2k}}^{-1}(C_{2k})$
$$
v=(x_{i_1},x_{j_1},\dots, x_{i_k},x_{j_k})
\mapsto
(t_{i_1},t_{j_1},\dots, t_{i_k},t_{j_k}), \ \text{\rm where} \
t_s=
\begin{cases}
(x_{j_{h-1}},x_{i_h},x_{j_{h}}) & \text{\rm for } s=i_h \\
(x_{i_{h}},x_{j_h},x_{i_{h+1}}) & \text{\rm for } s=j_h
\end{cases}
$$
is a bijection, where $h$ is considered modulo $k$.
Hence every $v\in \vartheta_{C_{2k}}^{-1}(C_{2k})$
can be uniquely decomposed into a sequence of triples. Since $v\in
\vartheta_{C_{2k}}^{-1}(C_{2k})$ there are exactly the following ten triples
$$
V_{D}=\{\textbf{ABA,ABD,BAB,BDB,BDC,DBD,DBA,DCD,CDC,CDB}\}
$$
and setting
$$
E_{D}=
\{\left((x_{j_{h-1}},x_{i_h},x_{j_{h}}),(x_{i_h},x_{j_{h}},x_{i_{h+1}})\right)
  \mid (x_{j_{h-1}},x_{i_h},x_{j_{h}})\in V_{D}\}
$$
we obtain the digraph ${D}$. Suppose we are given $v,v'\in
\vartheta_{C_{2k}}^{-1}(C_{2k})$ with $d(v,v')=1$ then we have the
following alternative {\small
$$
x_{j_{h-1}}=x_{j_h}={\bf B}:
\underbrace{\diagram
                       &  \fbox{{\bf D}}\ar@{-}[r]  &  {\bf B}  \\
{\bf B}\ar@{-}[ur] \ar@{-}[r]  & \fbox{{\bf A}}
\ar@/^2pc/@{->}[u]|{}
\ar@/_2pc/@{<-}[u]|{}
\ar@{-}[ur]  & \\
\enddiagram}_{(x_{j_{h-1}},x_{i_h},x_{j_{h}})}
\qquad x_{j_{h-1}}=x_{j_h}={\bf D}:
\qquad
\underbrace{\diagram
                                 &  \fbox{{\bf C}}\ar@{-}[r]  &  {\bf D}  \\
{\bf D}\ar@{-}[ur] \ar@{-}[r]  & \fbox{{\bf B}} \ar@{-}[ur]
\ar@/^2pc/@{->}[u]|{}
\ar@/_2pc/@{<-}[u]|{}
  & \\
\enddiagram}_{(x_{j_{h-1}},x_{i_h},x_{j_{h}})}
$$}

The idea is now to count all triples i.e.~$(x_{j_{h-1}},x_{i_h},
x_{j_{h}})$, $(x_{i_{h-1}},x_{j_{h-1}},x_{i_{h}})$ contained in
$\Theta=\{ {\bf BAB}, {\bf BDB}, {\bf DBD}, {\bf DCD}\}$ in
$\vartheta^{-1}_{C_{2k}}(C_{2k})$. Let next $R[x]$ be a polynomial
ring and $w\colon E_{D}\longrightarrow R[x]$ a function given by
$w(e)=x$ iff the arc $e$ has terminus $\tau\in\Theta$, otherwise
$w(e)=1$. If $\Gamma=e_{1}e_{2}\dots e_{\ell}$ is a walk of length
$\ell$ in $E_{D}$, then the weight of $\Gamma$ is defined by
$w(\Gamma)=w(e_1)w(e_2)\dots w(e_\ell)$. Introducing the formal
variable $x$ in $w$ allows us to count the triples in $\Theta$
within some $v\in\vartheta_{C_{2k}}^{-1}(C_{2k})$. The number of
closed walks of length $\ell$ in ${D}$ is $\sum_{v\in
{V_{D}}}{\left[{A_{D}}^\ell\right]}_{v,v} =\text{\sf
Tr}(A_{D}^\ell)$, where
$A_{D}$ is the adjacency matrix of ${D}$.\\
Suppose $B$ is a $p\times p$ matrix and $\{\eta_{i}\}_{i=1}^{p}$ are
all the eigenvalues of $B$, then we have ${\sf
det}B=\prod_{i}{\eta_{i}}.$ Let $\{\xi_{i}\}_{i=1}^{p}$ and
$\{\omega_{i}\}_{i=1}^{p}$ be all the eigenvalues of $I-yA$ and $A$
respectively, then we have $\xi_{i}=1-y\omega_{i}$, where $1\leq
i\leq p$. For the set of  all the nonzero eigenvalues of $A$,
$\{\omega_{i}\}_{i=1}^{r}$ we derive ${\sf det}(I-yA)=
\prod_{i=1}^{r}(1-y\omega_{i})$. We set $Q(y)={\sf det}(I-yA)$ and
have $p=10=\vert V_{D}\vert$, $A=A_{{D}}$ and $r=6$ for $x\neq 1$,
whence
\begin{equation}\label{E:reihe}
\sum_{\ell \ge 1} \text{\sf Tr}(A_{D}^\ell) y^\ell =\sum_{\ell \ge
1}(\omega_{1}^{\ell}+\dots+\omega_{r}^{\ell})y^{\ell}= \sum_{i=1}^r
\frac{\omega_iy}{1-\omega_iy}=\frac{-y\, Q'(y)}{Q(y)}.
\end{equation}
After some computation we derive
$Q(y)=1-2xy^2-x^2y^2+2x^{3}y^{4}-x^{4}y^{6}+2x^{3}y^{6}-x^{2}y^{6}-x^{2}y^{4}$
and the lemma follows from eq.~(\ref{E:reihe}). $\ \square$

{\bf Acknowledgments.}
We thank F.W.D.~Huang and L.C.~Zuo for helpful suggestions.
This work was supported by the 973 Project, the PCSIRT Project of the
Ministry of Education, the Ministry of Science and Technology, and
the National Science Foundation of China.

\bibliography{cm}
\bibliographystyle{plain}


\end{document}